\documentclass[a4paper]{article}
\usepackage{INTERSPEECH2021}
\usepackage{amsmath,graphicx}
\usepackage{multirow}
\usepackage{makecell}
\usepackage{graphicx,color}
\usepackage{comment}
\usepackage{bbm}
\usepackage{color,colortbl}
\usepackage{xcolor}
\usepackage{adjustbox}
\usepackage{enumitem}
\usepackage{subfigure}
\usepackage[linesnumbered, ruled, vlined, algo2e]{algorithm2e}
\usepackage{array}
\usepackage{caption}
\usepackage{soul}
\usepackage{makecell}
\usepackage{booktabs}
\usepackage[colorlinks=true,allcolors=blue]{hyperref}
\usepackage[numbers,sort&compress]{natbib}
\definecolor{LightBlue}{RGB}{233,243,250}

\newcommand{\eat}[1]{}

\usepackage{xspace}

\makeatletter
\DeclareRobustCommand\onedot{\futurelet\@let@token\@onedot}
\def\@onedot{\ifx\@let@token.\else.\null\fi\xspace}

\makeatother
\usepackage{amssymb}
\usepackage{amsmath,amsfonts}
\usepackage{amsopn,bm,mathtools}

\usepackage{nicefrac}       % compact symbols for 1/2, etc.
\newcommand{\vct}[1]{\boldsymbol{#1}} % vector
\newcommand{\ProbOpr}[1]{\mathbb{#1}}
\newcommand{\expect}[2]{%
\ifthenelse{\equal{#2}{}}{\ProbOpr{E}_{#1}}
{\ifthenelse{\equal{#1}{}}{\ProbOpr{E}\left[#2\right]}{\ProbOpr{E}_{#1}\left[#2\right]}}} % Expectation: syntax: E{1}{2} = E_1[2], E{}{2}=E[2], E{1}{} = E_1
\newcommand{\var}[2]{%
\ifthenelse{\equal{#2}{}}{\ProbOpr{VAR}_{#1}}
{\ifthenelse{\equal{#1}{}}{\ProbOpr{VAR}\left[#2\right]}{\ProbOpr{VAR}_{#1}\left[#2\right]}}} % Expectation: syntax: V{1}{2} = V_1[2], V{}{2}=V[2], V{1}{} = V_1
\newcommand{\vq}{\vct{q}}

\title{Unsupervised Data Selection via Discrete Speech Representation for ASR}
\name{Zhiyun Lu, Yongqiang Wang, Yu Zhang, Wei Han, Zhehuai Chen, Parisa Haghani}
\address{Google Inc.}
\email{\{zhiyunlu,yqw,ngyuzh,weihan,zhehuai,parisah\}@google.com}

\begin{document}

\maketitle
\begin{abstract}
Self-supervised learning of speech representations has achieved impressive results in improving automatic speech recognition (ASR). In this paper, we show that data selection is important for self-supervised learning. We propose a simple and effective unsupervised data selection method which selects acoustically similar speech to a target domain. It takes the discrete speech representation available in common self-supervised learning frameworks as input, and applies a contrastive data selection method on the discrete tokens.
Through extensive empirical studies we show that our proposed method reduces the amount of required pre-training data and improves the downstream ASR performance. 
Pre-training on a selected subset of 6\% of the general data pool results in 11.8\% relative improvements in LibriSpeech test-other compared to pre-training on the full set.
On Multilingual LibriSpeech French, German, and Spanish test sets, selecting 6\% data for pre-training reduces word error rate by more than 15\% relatively compared to the full set, and achieves competitive results compared to current state-of-the-art performances.
\end{abstract}

%On the LibriSpeech test-other set, it shows 11.8\% relative word error rate reduction by pre-training on 6\% of selected subset compared to pre-training on the full set. 

% token-based data selection method. 
% self-supervised learning of representations from raw audio data

%!TEX root = main.tex
\section{Introduction}
% Motivation
Self-supervised pre-training has demonstrated great success in learning representations from unlabeled data and improving the downstream automatic speech recognition (ASR) task~\cite{van2018representation,pascual2019learning,baevski2020wav2vec,baevski2019vq,chung2021w2v}. 
The paradigm learns good representations from unlabeled audio through a proxy task which predicts the masked part of input from its visible parts. Popular proxy tasks include contrastive task~\cite{van2018representation,schneider2019wav2vec};
BERT-style masked language modeling task~\cite{chung2021w2v,hsu2021hubert}; and reconstruction task~\cite{van2017neural}. 

While the majority of research in the field focus on designing better context prediction tasks and self-supervision losses, an important question is left unanswered: what \emph{data} should we use to do self-supervised pre-training? Is it always the more data the better? 
In supervised learning, it is well-known that learning from data of matched domain is important~\cite{seltzer2013investigation,likhomanenko2020rethinking}; in semi-supervised learning (teacher-student learning)~\cite{kahn2020self,park2020improved,doutre2021improving}, pseudo-label filtering and data weighting is carefully studied~\cite{charlet2001confidence,wessel2004unsupervised,vesely2017semi}. Few recent works study the effect of data selection on self-supervised learning ~\cite{kawakami2020learning,hsu2021robust,baskar2022ask2mask}.~\cite{hsu2021robust} shows that domain shift hurts the self-supervised pre-training, and~\cite{baskar2022ask2mask} shows that frame selection and data reweighting is helpful.

\begin{figure}[tb]
    \centering
     \includegraphics[width=0.45\textwidth]{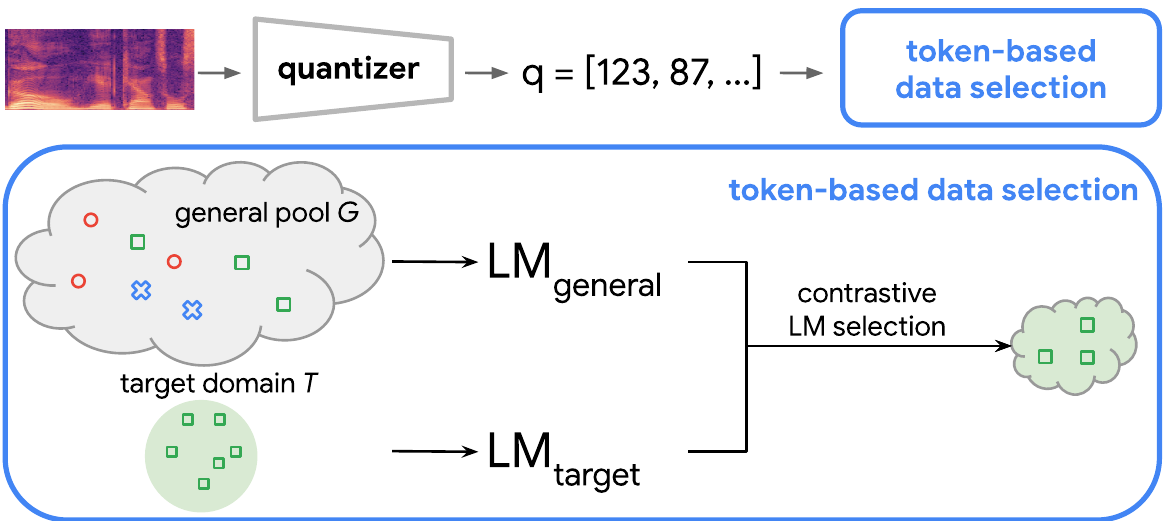}
    \caption{Unsupervised data selection framework: the utterance is first quantized to a sequence of discrete tokens, and then we apply token-based data selection method to the discrete speech sequences. The quantizer is readily available from popular self-supervised learning methods. We use contrastive language model selection as the selection method.} \label{fig:diagram}
\vspace{-0.2in}
\end{figure}

In this paper, we show that data selection is important for self-supervised learning. To this end, we propose a simple and flexible unsupervised data selection framework.
As shown in Fig.~\ref{fig:diagram}, we first encode an utterance into a discrete speech representation as is commonly done in self-supervised learning methods~\cite{baevski2020wav2vec,chung2021w2v}, and then apply a token-based data selection method on the discrete token sequences. The framework can work with different quantizers and data selection modules.
% ~\cite{chiu2022self} shows that discrete speech tokens are good targets in BERT-style self-supervised learning. 

Inspired by~\cite{chiu2022self}, we use the discrete speech tokens as the features for data selection. 
For the data selection module, in order to select data that is close to the target domain, we adopt the language model based contrastive method introduced in~\cite{moore2010intelligent}. Contrastive data selection is a well-known technique in the NLP and speech communities, and has been widely used in machine translation~\cite{axelrod2011domain,van2017dynamic}, and ASR~\cite{huang2022sentence,mezzoudj2018textual,chen2021injecting}. It computes a domain relevance score for each utterance and filters by a threshold to keep those closest to the target domain. 
% language modeling for ASR~\cite{mezzoudj2018textual} and text-to-speech for pre-training~\cite{chen2021injecting}

% this is not the only way, but
% \todo{Explain why discrete speech representation is a good feature for data selection.}
% from a large pool of non-domain-specific audio

%parisah: I rewrote this a bit to make it more concrete.
Our proposed data selection framework has the following benefits: \textit{i)} It improves the downstream ASR performance by exploiting similar non-domain-specific audio. \textit{ii)} It is data efficient and greatly reduces computation in pre-training as it can select a small subset of data for pre-training.  \textit{iii)} It is \emph{unsupervised}: it does not require labeled data which is appealing for low resource languages and domains. \textit{iv)} It does not require extra feature learning as it uses available discrete tokens from self-supervised learning frameworks. \textit{v)} Lastly, our experiment shows that it is not sensitive to the choice of quantizer nor other hyper-parameters thus almost tuning free.

 % to extract discrete tokens 

%The proposed data selection method improves the fine-tuning ASR performance as well as the data efficiency of pre-training. It greatly reduces the computation to exploit a large amount of non-domain-specific audio. Secondly, the method is \emph{unsupervised} and does not require any human supervision in the loop. This is in particular appealing for low resource languages and domains. Thirdly, it does not require extra feature learning by re-using the available discrete representation from the self-supervised learning framework. Lastly, it is not sensitive to the choice of quantizer to extract discrete tokens nor other hyper-parameters thus almost tuning free.
% since filtering a large amount of data is much  cheaper than pre-training on all the data. 
% 

We demonstrate the effectiveness of the data selection method for pre-training on LibriSpeech and multilingual LibriSpeech (MLS) datasets. By selecting 60k hours of YouTube speech which is acoustically similar to LibriSpeech for pre-training, we get word error rate (WER) 1.7\% and 3.0\% on test-clean and test-other set, after fine-tuning on LibriSpeech 960h. Our method reduces WER by 11.8\% relative on test-other compared to pre-training on the full set of 1 million hours. This performance is close to the state-of-the-art model pre-trained with the in-domain Libri-light 60k data~\cite{chung2021w2v}.  
On MLS French, German, and Spanish test sets, pre-training on selected 60k hours of YouTube speech gets WER of 3.7, 3.3, 3.2 respectively, which is better than the state-of-the-art WERs achieved by multilingual ASR model~\cite{bai2021joint}.
The relative WER reduction compared to pre-training on over 1 million hours is 18\%, 15\%, and 26\%. 
Our selection method is also applicable in supervised learning, as shown through experiments on WSJ and CHiME-6.

% We introduce the data selection method in \S~\ref{sec:method}, describe the experimental setup in \S~\ref{sec:setup}, and discuss the results in \S~\ref{sec:result}.

% The data selection method works by taking the discrete speech units from common self-supervised learning methods as input,

%!TEX root = main.tex
\section{Unsupervised Data Selection via Discrete Speech Representation}
\label{sec:method}
We first define the task of data selection for self-supervised learning and discuss related works in \S~\ref{sec:method_problem}, and then describe our unsupervised data selection method in \S~\ref{sec:method_token} and~\ref{sec:method_selection}. 

\subsection{Data selection for self-supervised learning}\label{sec:method_problem}
% Problem statement
Data selection improves ASR performance and data efficiency by identifying the most informative training examples. 
Concretely, the task of data selection for self-supervised learning is,
given labeled data from a target domain $T$ and \emph{unlabeled} data from a general pool $G$, selecting the best subset of $G$ for self-supervised pre-training, such that the model pre-trained on that subset and then fine-tuned on $T$ achieves the best performance for the target domain. 
We assume that $G$ is much larger than $T$. % and contains unlabeled data.

Confidence filtering is a classic method in data selection for ASR~\cite{zavaliagkos1998utilizing,chan2004improving}, and
has been successfully applied to end-to-end models~\cite{park2020improved,zhang2021bigssl,hwang2021large}. However, confidence methods focus on data of good \emph{transcript quality}, which might not be useful for self-supervised pre-training.
Moreover, training a confidence model requires supervised data, which is not often feasible. ~\cite{baskar2022ask2mask} proposes to do frame-level data selection and utterance-level reweighting, which is complementary to our task. 

We propose to do data selection by applying contrastive data selection on discrete speech representations. It selects data that is acoustically similar to the target domain, and does not require any labeled data in the loop.

% On the other hand, our proposed method addresses the audio domain closeness for data selection. On the contrary, our method does not require any labeled data.

\subsection{Discrete speech representation}\label{sec:method_token}
 % While the community is actively exploring new self-supervised objectives, 
An emergent trend in self-supervised learning for speech is to learn discrete tokens from the continuous speech signals~\cite{chorowski2019unsupervised,liu2020towards}. The discrete representations make it applicable to NLP methods that require discrete inputs, for example BERT-style pre-training algorithms. Empirically, it leads to better results compared to without quantization~\cite{baevski2019vq,baevski2020wav2vec}.
%  harness advances in NLP to speech % easy to apply NLP methods to speech,

% The quantizer module serves , and 
A quantizer maps continuous features into discrete tokens from a learnt codebook. It can be placed at different depths of the network, which leads to latent codes of different semantic level. In wav2vec 2.0 family of models~\cite{baevski2020wav2vec,chung2021w2v}, the quantization is immediately after the feature encoder and before any transformer/Conformer representation learning layers. Therefore the discrete tokens is of low semantic level, and could contain information like pitch, background noise, and other cofounding details of the audio signal. On the other hand, in the generative VQ-VAE, the quantization is at the top of all representation layers, where the discrete code is more abstract.
~\cite{chorowski2019unsupervised} shows the VQ-VAE quantized code is predictive of the phonetic content of the utterances.
Moreover, the discrete token is used differently in the training objectives. W2v-BERT uses the token ID as the target label in the cross-entropy loss of the masked language model task; VQ-VAE uses the discrete token to look up a 1-of-K embedding vector, which is then used to reconstruct the spectrogram. It's an open research question on how the quality of the quantization affects the self-supervised learning~\cite{chiu2022self}.
In this work, we experiment with both w2v-BERT and VQ-VAE discrete token sequences as input to the contrastive data selection module described next. 

% \todo{why is discrete speech representation a good feature for data selection.} 

\subsection{Contrastive data selection}\label{sec:method_selection}
Contrastive data selection selects examples from a source corpus that are matched to a target domain. It was originally proposed on text data~\cite{moore2010intelligent}, and has since been successfully applied in other domains such as machine translation and ASR. We extend this method to work on discrete speech tokens.

More specifically, we train two language models (LM) on corpora of the target domain $T$ and of general domain $G$ using the discrete speech tokens as input. For each utterance, we compute the log probability difference~\cite{moore2010intelligent,chen2021injecting} between the two LMs normalized by the number of tokens in the utterance. Assume $\vq$ is the vector quantized representation of the utterance, the domain relevance score is
% \begin{align*}
   $\frac{\log P_{T}(\vq) - \log P_{G}(\vq)}{\text{length}(\vq)}$,
% \end{align*}
where $P_{T}$ and $P_{G}$ are the probabilities of the target and general domain LM respectively.
We select utterances with the top domain relevance scores. See Fig.~\ref{fig:diagram} for an illustration.

\section{Experimental Setup}\label{sec:setup}

While the proposed selection method is unsupervised, it can be applied to both unlabeled data for pre-training, and labeled data for supervised learning. We investigate both settings in the empirical study. We describe the datasets in \S~\ref{sec:data}, the quantizer in \S~\ref{sec:quant} and contrastive selection setup in \S~\ref{sec:contrastive}. We provide model architecture in \S~\ref{sec:model}, and training hyper-parameters in \S~\ref{sec:training}. In each subsections, we present pre-training and supervised learning experiments 
separately.
% on both unlabeled data for pre-training and labeled data for supervised learning. 

\subsection{Data}\label{sec:data}
In the data selection for pre-training experiments, we use LibriSpeech and MLS as the target dataset, and YT-U as the general pool. We first pre-train the encoder on YT-U and then fine-tune on the LibriSpeech or MLS labeled set. The quantizer for speech tokenization is trained on YT-U.
In the data selection for supervised learning experiments, we use WSJ and CHiME-6 as the target datasets, and the People's Speech dataset as the general pool. We add the selected People's Speech data to the in-domain training set for supervised learning. We reuse a quantizer trained on SpeechStew for these experiments. See Table~\ref{tab:dataset} summarizes the datasets in each experiment.

\begin{table}[h]
\centering
\vspace{-0.1in}
    \caption{Summary of datasets used in the experiments.}     \label{tab:dataset}
    \vspace{-0.1in}
\begin{adjustbox}{max width=\columnwidth}
\begin{tabular}{lccc}
    \toprule
   Experiment & Target domain $T$  & General pool $G$ & Quantizer data \\ \midrule 
   Pre-train (\S~\ref{sec:librispeech},\ref{sec:ablation}) & LibriSpeech / MLS  & YT-U & YT-U \\
   Supervised (\S~\ref{sec:supervised})  & WSJ / CHiME-6 & People's Speech & SpeechStew \\
      \bottomrule
\end{tabular}
\end{adjustbox} 
\end{table}

%!TEX root = main.tex
\begin{table*}[tb]
    \centering
    \caption{Comparing WERs on Librispeech dev and test sets when the encoder is pre-trained on YouTube with and without data selection. The model is fine-tuned on 960h and 100h. Pre-training on 6\% of selected subset reduces WER by 11.8\% relative on test-other set.}
    \label{tab:librispeech}
    \vspace{-0.1in}
    \begin{tabular}{lrcccc}  
\multirow{2}{*}{Pre-train data}  & Pre-train & \multicolumn{2}{c}{Fine-tune 960h} & \multicolumn{2}{c}{Fine-tune 100h}  \\
\cmidrule(lr){3-4}  \cmidrule(lr){5-6} 
& \# hours & Dev clean / other & Test clean / other & Dev clean / other & Test clean / other \\
\toprule
Libri-light~\cite{chung2021w2v} &  60k   & 1.5 / 2.9 & 1.5 / 2.9  & 2.4 / 4.4 & 2.5 / 4.6   \\ \midrule
YT-U-En & 1,000k           & 1.7 / 3.3 & 1.8 / 3.4 & 3.0 / 5.8 & 3.1 / 5.9   \\ 
YT-U-En select & 60k   & 1.6 / 2.9 &  1.7 / 3.0 & 2.8 / 5.0 & 2.8 / 5.2 \\  % \midrule
YT-U-En select + Libri-light & 120k   & 1.5 / \bf{2.7} &  1.6 / \bf{2.8} & 2.4 / \bf{3.9} & 2.5 / \bf{4.4} \\ 
    \bottomrule
    \end{tabular}
    \vspace{-0.1in}
\end{table*}

{\noindent \bf{LibriSpeech}}~\cite{panayotov2015librispeech}
We consider both LibriSpeech 100 hours and 960 hours as the supervised fine-tuning data. We report word error rates (WERs) on the dev-clean, dev-other, test-clean, and test-other evaluation sets. We also use Libri-light 60k in pre-training as a baseline for comparison.

{\noindent \bf{Multilingual LibriSpeech} (MLS)}~\cite{pratap2020mls}
We use French, German, and Spanish sets from MLS corpus. We use MLS-full as the fine-tuning supervised data, and report WERs on test sets.

{\noindent \bf{YouTube unsupervised data (YT-U)}} is collected from speech-heavy videos that cover many different domains, including lectures, news, and etc. The raw audio is segmented by a voice activity detector~\cite{zazo2016feature} to a length of 32 seconds per utterance. We prepare speech data of 1.0 million hours in English (YT-U-En), 1.2 million in French (YT-U-Fr), 1.1 million in German (YT-U-De), and 1.0 million in Spanish (YT-U-Es). We use YT-U to refer to one of YT-U-En, YT-U-Fr, YT-U-De, and YT-U-Es, when it is clear from the context.
% We use YT-U in both pre-training and quantizers training. % Please refer to~\cite{zhang2021bigssl} for more details of the data. 
% a total number of 1.0 million hours
% We use  American English, French, German, and Spanish.

{\noindent \bf{Wall Street Journal (WSJ)} }  (LDC93S6B, LDC94S13B) is 80 hours of read speech from Wall Street Journal news text corpus. We report WERs on the eval92 test set, and score our results with the Kaldi script~\cite{povey2011kaldi}.

{\noindent \bf{CHiME-6}}~\cite{watanabe2020chime} is a set of approximately 40 hours of noisy distant microphone conversational speech in everyday home environments. We use the official front-end enhancement recipe to augment the dataset, and report WER on the test set where guided source separation
with 12 channels enhancement is used. 

{\noindent \bf{People's Speech}}~\cite{galvez2021people} is a public speech dataset from diverse sources including movies, TV, local news, and etc. We use the 17.1k hours clean subset of People's Speech. 

{\noindent \bf{SpeechStew}}~\cite{chan2021speechstew} is an ensemble dataset combining 7 public speech corpora, including AMI, Common
Voice, English Broadcast News, LibriSpeech, Switchboard/Fisher, TED-LIUM v3, and WSJ. We use the unlabeled speech in SpeechStew to train the quantizer.

% ~\cite{carletta2005ami} ~\cite{ardila2019common} (LDC97S44, LDC97T22, LDC98S71 and LDC98T28) ~\cite{panayotov2015librispeech} (LDC2004T19, LDC2005T19, LDC2004S13, LDC2005S13 and LDC97S62)

\subsection{Quantizer}\label{sec:quant}
In the pre-training experiment, we use both the w2v-BERT~\cite{chung2021w2v} and the VQ-VAE~\cite{van2017neural} as the quantizer. In the supervised learning experiment, we use a VQ-VAE quantizer. In w2v-BERT quantizer, the codebook vocab size is 1024 and the codebook dimension is 1024. In VQ-VAE, the vocab size is 8192 and the dimension is 16. For both models, we apply Gumbel-Softmax technique~\cite{jang2016categorical} to the quantization operation, which makes the argmax differentiable through a temperature parameter in the backward pass. The temperature is gradually annealed towards a small but non-zero value during training.

\subsection{Contrastive data selection}\label{sec:contrastive}
We use the same set of contrastive data selection parameters across all experiments.
We build $N$-gram language model (LM) with $N=5$ and back-off. We use the Kneser-Ney LM interpolation algorithm~\cite{heafield-etal-2013-scalable} to estimate LM probabilities. We use a few hundred to a few thousand hours of data as input. In the pre-training experiments, we use the in-domain training set for the target LM, and a random subsample of a few hundred hours YT-U for the general LM. In the supervised experiment, we use WSJ or CHiME-6 training set to build the target LM, and the People's Speech clean set to build the general LM. As we will show in the experiment, the method is not sensitive to the amount of data in LM training.

\subsection{Model architecture}\label{sec:model}
In all our experiments, we use 80-dimensional log-mel filter bank coefficients as the acoustic inputs, computed with a 25ms window and shifted every 10ms. For transcript tokenization, we use a 1024-token WordPiece model~\cite{schuster2012japanese} for all English experiments, and 4096-token WordPiece models for French, German, and Spanish experiments.
%  constructed from the transcripts of the LibriSpeech 960 hours

In the pre-training experiments, we follow the w2v-BERT XL setup in~\cite{chung2021w2v} which is a 24 layer Conformer. The feature encoder 
has two 2D-convolution layers with strides (2,2).
%In the pre-training experiments, we use a 24 layer Conformer model~\cite{gulati2020conformer} and follow the w2v-BERT XL setup in~\cite{chung2021w2v} to feed 
We use the 12-th Conformer block hidden features to compute the contrastive loss and the last (24-th) layer features to compute the masked prediction loss.
%in w2v-BERT pre-training. It is the same as w2v-BERT XL in~\cite{chung2021w2v}. Please refer to~\cite{chung2021w2v} for more details.
We use a RNN-T model for fine-tuning. The decoder is a two-layer LSTM with a hidden dimension of 640. The model has 600 million parameters. 
In the supervised learning experiments, we use a 17 layer Conformer model of 119 million parameters, the same as ConformerL in~\cite{gulati2020conformer}.

\subsection{Training details}\label{sec:training}
In the pre-training experiments, we use w2v-BERT as the pre-training recipe. We use a masking length of 400ms with masking probability of 0.065. We use Adam optimizer~\cite{diederik2014adam}, and a transformer learning rate schedule~\cite{vaswani2017attention} with 1e-3 peak learning rate and 25,000 warm-up steps. The batch size is 4096.
In the fine-tuning, we use different learning schedules for the pre-trained encoder and the decoder. The encoder has a peak learning rate of 3e-4 and 5,000 warm-up steps, while decoder has 1e-3 and 1,500 steps. The batch size is 256.

For the supervised learning experiment, we use Adam optimizer with a chedule of 0.002 peak learning rate and 10,000 warm-up steps. The batch size is 256 for WSJ experiment, and 2048 for CHiME-6 experiment. We find a larger batch size is essential for CHiME-6, otherwise the model may fail to train.
% We find that a smaller batch size gives better result on WSJ. On the other hand, a larger batch size is essential for CHiME-6, otherwise the model might fail to train.

% ################## 
% remove part
% ##################

% \begin{table}[tb]
% \centering
%     \caption{Summary of datasets used in the experiments.}

% \begin{adjustbox}{max width=\columnwidth}
% \begin{tabular}{ccc|ccc}
%     \toprule
%     & \multicolumn{2}{c|}{Unlabeled} & \multicolumn{3}{c}{Labeled} \\ 
%       Dataset  & YT-U  &  Libri-light & YT-L & LibriSpeech & AMI (ihm / sdm) \\ % 
%       \# hours & 1m & 600k  & 600k & 960 & 100 (? / ?) \\ 
%       \bottomrule
% \end{tabular}
% \end{adjustbox}
%     \label{tab:dataset}
% \end{table}

% The two models thus the quantizers are trained with different losses, contrastive and masked language model losses for w2v-BERT and reconstruction loss for VQ-VAE. The quantizers are also placed at different depths in the model, thus the discrete units are likely

% Quantized representations are also learnt with different objectives in different models. In w2v-BERT, it is trained with contrastive loss and the mlm loss, while variational lower bound in VQ-VAE. 
%!TEX root = main.tex
\section{Experimental Results}\label{sec:result}
We present key results of data selection for pre-training in \S~\ref{sec:librispeech}, followed by ablation studies in \S~\ref{sec:ablation}. Then we show the selection method is also applicable to supervised learning in \S~\ref{sec:supervised}. % Lastly we discuss in \S~\ref{sec:discuss}.

\subsection{Data selection for self-supervised pre-training}\label{sec:librispeech}
% We report WERs after fine-tuning to demonstrate the effectiveness of data selection in self-supervised pre-training. 

\begin{table}[tb]
    \centering
    \caption{Comparing WERs on MLS test sets in French, German and Spanish when the pre-training data is YouTube with and without data selection. The model is fine-tuned on MLS-Full. Pre-training on the selected subset reduces WERs by more than 15\% relative in all 3 languages. Our best WER is better than the state-of-the-art performance obtained by multilingual models.}
    \label{tab:mls}
    \vspace{-0.05in}
    \begin{tabular}{l@{\hskip -0.05pt}rccc}   % @{\hskip -0.05pt}
Pre-train data & \# hrs & French &  German & Spanish  \\ 
\toprule
\multicolumn{5}{l}{{\bf{Prior work}} \ \ \ multilingual model} \\
SoTA & & 4.9~\cite{li2021scaling} & 4.1~\cite{bai2021joint} & 3.7~\cite{bai2021joint}  \\
\midrule
\multicolumn{5}{l}{{\bf{Our work}}  \ \ \ \ \ monolingual model}  \\
% No pre-train & 0 & \\
YT-U  &  1,000k &  4.5 &  3.9 & 4.3  \\ 
YT-U select  & 60k & \bf{3.7} &  \bf{3.3} & \bf{3.2}  \\ 
% YT-U-{lang} &  1m  &   5.4 / 4.5   &  3.3 / 3.9 & 3.0 / 4.3 \\ 
% YT-U-{lang} select & 60k & 4.3 / 3.7 & 2.9 / 3.3 & 3.4 / 3.2 \\ 
    \bottomrule
    \end{tabular}
    \vspace{-0.2in}
\end{table}

\begin{table}[tb]
    \centering
    \caption{Our unsupervised data selection method outperforms random sample and confidence filtering baselines. Both confidence and our method are better than no data selection.}
    \label{tab:baseline}
    % \begin{adjustbox}{width=0.47 \textwidth}
    \begin{tabular}{l@{\hskip -0.03pt}rcc}  
Selection & \# hours & Dev clean / other & Test clean / other \\ 
\toprule
no selection & 1,000k & 1.9 / 3.7 & 2.0 / 3.9 \\
random  & 60k & 1.9 / 4.1 &  2.1 / 4.1 \\ 
% random & \\
confidence & 60k & 1.7 / 3.3 & 1.9 / 3.5 \\
% ours (500k steps) & \bf{1.6} / \bf{2.9} &  \bf{1.7} / \bf{3.0} \\ 
ours  & 60k & 1.7 / \bf{3.2} & \bf{1.8} / \bf{3.3} \\
\bottomrule
\end{tabular}
\vspace{-0.1in}
% \end{adjustbox}
\end{table}
 
Table~\ref{tab:librispeech} compares WERs on LibriSpeech dev and test sets when the models are pre-trained on YT-U-En with and without data selection, 
and then fine-tuned on LibriSpeech 960h or 100h. 
We compare with state-of-the-art performance from~\cite{chung2021w2v} in the first row, where the pre-training data is the in-domain Libri-light 60k hours. For our models, we pre-train the encoder for up to 800k steps. Since there is no golden rule in picking pre-training checkpoints, we fine-tune at every 100k checkpoints. We pick the best one in terms of fine-tuning WERs on dev sets for each method to report. 
Comparing the YT-U with YT-U select row, we reduce WER by 5\% and 11\% relative on the test clean and other set respectively, and reduce the amount of pre-training data to 6\%. In the last row, by combining the selected YT-U data with Libri-light, we are able to slightly improve the state-of-the-art WER on the test-other set.

Table~\ref{tab:mls} shows WERs on MLS French, German, Spanish test sets. For our monolingual models, the pre-training data is YT-U and the fine-tuning data is MLS-full. To the best of our knowledge, the state-of-the-art (SoTA) WERs on these test sets are achieved by multilingual models in~\cite{bai2021joint,li2021scaling}, as detailed in the first row. We pre-train for 100k steps for fast experimentation. %, as the relative comparison stays the same when we pre-train for more steps.
Comparing YT-U with YT-U select, the WER reduction is 18\%, 15\%, and 26\% for French, German and Spanish. And the amount of data is less than 6\%.

% We also provide a monolingual model baseline that is trained from scratch on MLS-full for reference.

In Table~\ref{tab:baseline} we compare the proposed selection method with random subsample and confidence filtering. We provide WERs with no data selection for reference. For the confidence filtering method, we use a strong RNN-T model of 183 million parameters trained from around 150k hours of labeled YouTube data. We interpret RNN-T loss as the confidence measure following~\cite{zhang2021bigssl}. 
All methods pre-train for 100k steps and fine-tune on 960h. Both confidence filtering and our method are better than no data selection. Note the confidence method requires labeled YouTube data. In contrast, our method does not require any paired data, yet still outperforms the confidence method. 

% \ie in-domain labeled data w.r.t. the data pool.
% \eat{ on the same amount of 60k hours speech}
% We pre-train all variants for 100k steps, as the relative comparison stays the same when we pre-train for more steps. 

\subsection{Ablation studies}\label{sec:ablation}
We conduct ablation studies on the choice of the quantizer, and hyper-parameters of the contrastive selection method. For all variants in this section, the pre-training data is 60k select YT-U-En, and the fine-tuning data is LibriSpeech 960 hours. We pre-train for 100k steps, as the relative comparison stays the same for more steps.

Table~\ref{tab:quantizer} compares WERs when different quantizers are used to extract the discrete tokens for data selection. Both w2v-BERT and VQ-VAE quantizers perform similarly. The proposed method is not sensitive to the choice of the quantizer. 

\begin{table}[tb]
    \centering
    \caption{The effect of quantizer on fine-tuning WERs. The pre-training data is YT-U selected with w2v-BERT or VQ-VAE quantizers. Both quantizers perform similarly.}
    \label{tab:quantizer}
    \begin{tabular}{lcc}  
Quantizer & Dev clean / other & Test clean / other \\ 
\toprule
w2v-BERT  &   1.7 / 3.2 & 1.8 / 3.3 \\ 
VQ-VAE  &  1.7 / 3.3 & 1.8 / 3.4 \\ 
% w2v-BERT (500k)  &  1.6 / 2.9 &  1.7 / 3.0 \\ 
% VQ-VAE (500k) & 1.7 / 3.0 &  1.7 / 3.1 \\ 
\bottomrule
    \end{tabular}
    \vspace{-0.1in}
\end{table}

Table~\ref{tab:lm_data} compares WERs when different amount of data is used for LM training in the contrastive selection module. The quantizer is w2v-BERT. Our method is not sensitive to the amount of data used in LM training, once in a reasonable range.
% The pre-training data is 60k YT-U speech selected using scores computed from different LMs. The LM training data is detailed in the first column. 
% As shown in Table~\ref{tab:lm_data}, 

\begin{table}[tb]
    \centering
    \caption{The effect of LM training data on fine-tuning WERs. The pre-training data is YT-U selected with LMs trained from different data. All variants perform similarly.}
    \label{tab:lm_data}
    \begin{tabular}{lccc}  
LM data \# hours & Dev clean / other & Test clean / other \\ 
\toprule
960h / 100h & 1.7 / 3.2 & 1.8 / 3.3 \\
100h / 100h & 1.6 / 3.2 & 1.8 / 3.4 \\ 
960h / 1 million h   & 1.7 / 3.1 & 1.7 / 3.4 \\ 
\bottomrule
    \end{tabular}
\end{table}

\subsection{Data selection for supervised learning}\label{sec:supervised}
While our work is mainly motivated to perform data selection for pre-training, the proposed method is also applicable to labeled data. We can select transcribed speech close to the target domain and add it to the training set to improve the ASR performance. This is useful for low resource domains. 

% we add utterances from People's Speech dataset to the original in-domain training set in ASR training.
% WSJ and CHiME-6 training sets have 80 and 40 hours of speech respectively. 

To demonstrate the performance in this setting, we apply data selection to the People's Speech dataset to improve ASR performance on WSJ and CHiME-6. We select around 800 hours of data from People's Speech and add that to the in-domain training data for supervised learning. 
In Table~\ref{tab:supervised}, we compare three data selection methods: random sample, text contrastive LM selection, and our unsupervised method. We do not include confidence method because the ground-truth transcript is already given.
We also present WERs when we use the original training set only from~\cite{chan2021speechstew}, and when we add the full 11.8k hours People's Speech data.

Adding extra labeled data from People's Speech greatly improves WERs on WSJ and CHiME-6. Our selection method outperforms random sample and text contrastive selection method. It reduce the amount of labeled data to 7\%, and the WER is 0.1\% worse compared to using the full People's Speech data. This result shows that we can apply the proposed method to select unlabeled data to send for transcription, and achieves good WER with small amount of annotation.

\begin{table}[tb]
    \centering
    \caption{WERs on WSJ and Chime6. We select 0.8k hours of data from People's Speech using different methods and add it to the training set. Our method outperforms other baselines. By using only 7\% of the selected subset, our method achieves close WER compared to the full set.}
    \label{tab:supervised}
    \vspace{-0.1in}
    \begin{adjustbox}{width=0.48\textwidth}
    \begin{tabular}{rl@{\hskip -0.1pt}rcc}  
\multicolumn{2}{c}{Training set} &  + \# hours & WSJ & CHiME-6 \\ 
\toprule
\multicolumn{2}{l}{in-domain \ \ wsj 80h / chime 40h ~\cite{chan2021speechstew}}  & 0 &  28.2 & 66.7 \\ 
\midrule
\multirow{3}{*}{in-domain} & 
+ random sample & 0.8k  & 3.1 &  51.2 \\  % 48.3
& + text contrastive &  0.8k &  3.1 & 49.7  \\ 
& + ours  & 0.8k & 2.7  &   48.0 \\     % \bf{46.5}
\midrule
in-domain & + full People's Speech  & 11.8k & 2.6 & 47.9 \\    %  49.3
    \bottomrule
    \end{tabular}
\end{adjustbox}
    \vspace{-0.2in}
\end{table}

Lastly, we compare the contrastive score ranking from quantized codes with contrastive score ranking from text. There is almost zero rank correlation between the two ordinal variables, measured by Kendall $\tau$ coefficient. It suggests that the two data selection methods can be complementary, and we leave empirical studies on this as future work.

\section{Conclusion}
We show that data selection is important for self-supervised pre-training. We propose a simple and effective unsupervised data selection method, which applies contrastive data selection on discrete speech representations. Our experimental results demonstrate the effectiveness of the proposed approach in improving the ASR performance in both unsupervised and supervised settings.
% when applied in unsupervised and supervised settings.
%that applying the proposed selection method in pre-training greatly improves downstream ASR performance. % It also has applicability in the supervised setting. 

\section{Acknowledgements}
We are grateful to Chung-Cheng Chiu, Pedro Moreno Mengibar, Neeraj Gaur and Trevor Strohman for their help and suggestions.

% \clearpage
% \newpage

\bibliographystyle{IEEEtran}
\section{References}
\renewcommand{\section}[2]{}
\setlength{\bibsep}{3pt plus 0.3ex}

\bibliography{ref}

\end{document}